\documentclass[11pt]{article}

\usepackage{amsfonts}
\usepackage{epsfig}
\usepackage{latexsym}
\usepackage{amsmath}
\usepackage{amssymb}
\usepackage{float}

\makeatletter
\@addtoreset{equation}{section}
\makeatother

\def\bequ{\begin{equation}}
\def\eequ{\end{equation}}
\def\barr{\begin{array}}
\def\earr{\end{array}}
\def\half{{1\over 2}}
\def\ben{\begin{equation}}
\def\een{\end{equation}}
\def\bena{\begin{eqnarray}}
\def\eena{\end{eqnarray}}

\def\b1{e^0}

\newcommand{\be}{\begin{equation}}
\newcommand{\ee}{\end{equation}}
\def\bea{\begin{eqnarray}}
\def\eea{\end{eqnarray}}

\def\nn{\nonumber}

\def\half {{1 \over 2}}

\def\be{\begin{equation}}
\def\ee{\end{equation}}
\def\bea{\begin{eqnarray}}
\def\eea{\end{eqnarray}}

\def\ft#1#2{{\textstyle{\frac{\scriptstyle #1}{\scriptstyle #2} } }}
\def\fft#1#2{{\frac{#1}{#2}}}

\def\bm{\bibitem}
\def\ep{{\epsilon}}

\def\lesssim{\mathrel{\hbox{\rlap{\hbox{\lower4pt\hbox{$\sim$}}}\hbox{$<$}}}}
\def\gtrsim{\mathrel{\hbox{\rlap{\hbox{\lower4pt\hbox{$\sim$}}}\hbox{$>$}}}}

\newcommand{\hoch}[1]{$\, ^{#1}$}

\newcommand{\auth}{\Large\bf{M. Cveti\v c\hoch{*}, G.W. Gibbons\hoch{\dagger}
and C.N. Pope\hoch{\ddagger,\dagger}}}

\begin{document}

\begin{flushright}
\hfill {UPR-1227-T\ \ \ 
DAMTP-2011-29\ \ \
MIFPA-11-13\ \ \
}\\
\end{flushright}

\begin{center}

{\Large{\bf More about Birkhoff's Invariant and Thorne's Hoop Conjecture
for Horizons}}

\vspace{30pt}
\auth

\large

\vspace{30pt}{\hoch{*}\it Department of Physics and Astronomy,\\
University of Pennsylvania, Philadelphia, PA 19104, USA}

\vspace{10pt}{\hoch{*}\it Center for Applied Mathematics and 
Theoretical Physics,\\
University of Maribor, Maribor, Slovenia}

\vspace{10pt}{\hoch{\dagger}\it DAMTP, Centre for Mathematical Sciences,\\
 Cambridge University, Wilberforce Road, Cambridge CB3 OWA, UK}

\vspace{10pt}{\hoch{\ddagger}\it George P. \& Cynthia W. Mitchell
Institute for\\ 
Fundamental Physics and Astronomy,\\ Texas A\& M University,
College Station, TX 77843-4242, USA}

\vspace{30pt}

\begin{abstract}

\rm

A recent precise formulation of the hoop conjecture in four spacetime
dimensions
is that the Birkhoff  invariant $\beta$ (the 
 least maximal length of any sweepout or foliation by circles)
 of an apparent horizon of energy $E$ and area $A$ 
should  satisfy $\beta \le 4 \pi E$. This conjecture together with the 
Cosmic Censorship or Isoperimetric inequality implies that
the length $\ell$ of 
the shortest non-trivial closed geodesic satisfies $\ell^2 \le \pi A$. 
We have tested these conjectures
on the    horizons of  all four-charged rotating black hole solutions of 
ungauged supergravity theories
and find that they  always hold. 
They  continue  to hold in the the presence of a negative cosmological
constant, and for multi-charged rotating solutions in gauged supergravity.
Surprisingly, they also hold for the Ernst-Wild static black holes
immersed in a magnetic field, which are asymptotic to the  Melvin solution.
In five spacetime dimensions we define $\beta$ as the least maximal area
of all sweepouts of the horizon by two-dimensional tori, and find
in all cases examined that  $ \beta(g) \le  \frac{16 \pi}{3} E$,
which we conjecture holds quiet generally for apparent horizons.
In even spacetime dimensions $D=2N+2$,
we find that for sweepouts by the product $S^1 \times S^{D-4}$,
$\beta$ is bounded from above by a certain dimension-dependent multiple
of the energy $E$.
We also find that $\ell^{D-2}$ is bounded from above by a certain 
dimension-dependent multiple of the horizon area $A$. 
Finally, we show that $\ell^{D-3}$ is bounded from above by a certain 
dimension-dependent multiple of the energy, for all Kerr-AdS black holes.

\end{abstract}

\end{center}

\pagebreak

\tableofcontents

\section{Introduction}

   Many years ago Thorne conjectured \cite{Thorne} that in four
spacetime dimensions 
\begin{quote} {\it Horizons form 
when and only when a mass $E$ gets
compacted into a region whose circumference in EVERY direction is
$C\le 4 \pi E$. }\end{quote} Since that time there has been a
 great deal of work making the idea more precise, and attempting to
establish its correctness or otherwise 
(see e.g. \cite{Yau,Tod,Senovilla}).\footnote{Indeed, as we discuss in
the appendix, one interpretation of Thorne's original statement of the 
conjecture appears to be violated by black holes in external magnetic 
fields.} 
Since then the hoop conjecture has been invoked 
in numerical relativity (see e.g. \cite{Ida1,Choptuik}) and
studies of  hole scattering in four and higher dimensions
 \cite{Eardley,Barrabes,Yamada,Ida2,Yoshino,Yoo}. It has also been
 suggested
that the hoop conjecture may provide a route to a precise formulation of the
idea that there is a minimal length in quantum gravity \cite{Basu}.     

  To begin with one needs  a definition of the total energy $E$.
One obvious possibility, in the asymptotically flat case that
Thorne had in mind,  is to take  
the ADM mass.    In order to define the circumference one needs
a notion of a surface that surrounds the matter. 
Thus one is  led to consider a Cauchy
surface $\Sigma$ containing an outermost marginally trapped surface or
``apparent horizon''  $S$
with induced metric $g$, and to assign to the pair $\{S,g\}$ 
a hoop radius $R$ or circumference $C=2\pi R$. 
In a recent note \cite{Gibbons} it has been
suggested that for topologically-spherical apparent horizons with metric $g$ 
in four-dimensional    
spacetimes one may take for $C$ the Birkhoff invariant $\beta(g)$, 
and so we propose\footnote{For this and all subsequent conjectures, we
assume that the dominant energy condition holds.}

\medskip
\noindent
{\bf Conjecture 1}: {\it The Birkhoff invariant $\beta(g)$ and the energy
$E$ of an apparent horizon, in $3+1$ dimensions, satisfy}
\be
\boxed{\beta(g) \le 4 \pi E\,.}\label{conj1}
\ee
Some evidence for 
conjecture 1 was presented in \cite{Gibbons}.\footnote{For the purposes
of the present work we are only interested in the necessity
of this  proposed inequality, and moreover we shall not discuss to what
extent it captures all of what Thorne had in mind when he made his
original conjecture. We shall also not be concerned with the question
of whether the ADM mass may be replaced by some quasi-local notion of mass.}
In the appendix we shall give some further discussion on the appropriateness
of taking $\beta(g)$ as the definition of the circumference, or hoop radius.
The bulk of our paper is concerned with whether or not conjecture 1, 
and related inequalities in four and higher dimensions, are
valid.

The Birkhoff invariant is defined as follows. 
If the matter obeys the dominant energy condition (which it does for ungauged
supergravity),
we can assume in four spacetime dimensions 
 that the apparent horizon is topologically 
spherical \cite{Hawking1,Gibbons1,Gibbons2,Hawking2}.
Now, suppose that $S=\{S^2,g \}$ is a sphere 
with arbitrary metric $g$ and $f:S \rightarrow {\Bbb R}$ is a function on
$S$ with just two critical points, a maximum and a minimum.
Each level set $f^{-1}(c),\,c \in {\Bbb R}$, has a length $\ell(c)$,
and for any given function $f$ we may define 
\ben
\beta (g;f) = {\rm max}_c\, \ell(c) \,.\label{betagf}
\een
(For example, for the ordinary unit sphere with spherical polar coordinates 
$(\theta,\phi)$, we may take $f=\cos\theta$ and 
$\ell(\cos\theta)=2\pi\sin\theta$.  Thus $\beta(g;\cos\theta)=2\pi$.)
We now define the Birkhoff invariant $\beta(S,g)$
by minimizing $\beta(g;f)$ over all such  functions,
\ben
\beta(g) = \inf _f \beta (g;f) \,.
\een
The intuitive meaning of $\beta(g)$ is the least length 
 of a closed flexible hoop
that may be slipped over the surface
$S$. To understand why, note that each function
$f$ gives a foliation of $S$ by a one-parameter family of 
simple closed curves $f=c$ which 
we may think of as the hoop at each 
``moment of  time''  $c$.    
$\beta(g;f)$ is the greatest length of the hoop during this process.
If we change the foliation we can hope to reduce  this greatest length,
and the infinum is the best that we can do. 
The phrase  ``moment of  time''  is in quotation marks
because we are not regarding $f$ as a physical  time function, 
but merely as a convenient  way of thinking about the geometry of $S$.
 
Clearly the definition of $\beta(g)$ does not depend upon
the spacetime's being asymptotically flat. Thus one is led to
conjecture that it continues to hold for asymptotically-AdS spacetimes,
with the ADM mass being replaced by the Abbott-Deser mass.
Another possibility is to consider a static black hole
immersed in an asymptotically Melvin magnetic field,
for which an appropriate notion of total energy is available. 
In section 2
of this paper we shall confirm  this  conjecture
for all the exact stationary black hole solutions
known  to us.
Note that to confirm the conjecture it suffices to  
bound $\beta(g;f)$ from above by $4 \pi E$ for some particular, 
conveniently chosen,
foliation $f$. We do not need to calculate $\beta(g)$ itself.  

It was shown by Birkhoff \cite{Birkhoff} that there is at least one
closed geodesic $\gamma$ on $S$ with length $\ell(\gamma)= \beta(g)$.
It follows that if $\ell(g)$ is the length of the shortest
non-trivial closed geodesic on $S$, then 
\ben
\ell(g) \le \beta (g) 
\een  
and so, if our conjecture is correct, it should be the case that

\medskip
\noindent
{\bf Conjecture 2}: {\it The length $\ell(g)$ of the shortest geodesic
and the energy
$E$ of an apparent horizon, in $3+1$ dimensions, satisfy}
\ben \boxed{
\ell(g) \le 4 \pi E \,. \label{conj2}} 
\een   
Again, to confirm conjecture 2 it suffices  to bound
$\ell(\gamma)$ by $4 \pi E$ for some particular, conveniently chosen,
geodesic $\gamma$. We do not need to calculate $\ell(g)$ itself. 
The simplest case in which this can be done is if $\{S,g\}$ admits
a fixed-point free  isometric action of ${\Bbb Z}_2$, an ``antipodal map.''
One may then pass to $S/{\Bbb Z}_2  \equiv{\Bbb R}{\Bbb P}^2$.
Since $\pi_1({\Bbb R}{\Bbb P}^2) = {\Bbb Z}_2$, there must be at least
one  closed geodesic $\gamma$ in this homotopy class, which may be obtained by
minimizing the length amongst all  non-trivial curves in this class.
To obtain an upper bound for $\ell(\gamma)$, it suffices to find an upper
bound for the distance between a point and its antipode.

   In fact Pu \cite{Pu} has shown in this case that if $A(g)$ is the 
area of $S$ then
\ben \boxed{
\ell(g) \le   \sqrt{\pi  A(g)} \,. \label{pu} } 
\een
However, the Penrose  inequality \cite{penrose} states that
\ben
\sqrt {\pi A} \le 4 \pi E\,,
\een  
and so the conjecture (\ref{conj2}) holds for those apparent horizons
$\{S,g \}$ 
admitting an antipodal map \cite{Gibbons}. In fact all event horizons
of regular black holes solutions 
known to us in four spacetime dimensions admit an antipodal map
and thus satisfy (\ref{conj2}). 

   Given the current interest in higher dimensions, it is natural
to attempt to extend these conjectures beyond four  dimensions,
and then to test them against  known exact solutions. 
This we do in section 3 of the present paper,
for most of the exact five-dimensional black holes solutions known to us
that have as horizon a topological 3-sphere. In general, these have
$\{S,g\} \equiv \{S^3,g\} $ for which $g$ is not the round 3-sphere metric.
In the cases that we study it is invariant under the action of 
$U(1)\times U(1)$.  Thus   Birkhoff's
invariant is obtained by considering the area of the leaves 
of a Clifford type foliation of  $S^3$ by 2-tori $S^1 \times S^1$ with two
singular linked $S^1$ leaves. For a review of mathematical results
on such higher-dimensional ``sweepouts,'' the reader may consult 
\cite{ColdingLellis}.   We propose

\medskip
\noindent
{\bf Conjecture 3}: {\it The Birkhoff invariant $\beta(g)$ for $S^1\times S^1$ 
sweepouts, and the energy
$E$ of an apparent horizon, in $4+1$ dimensions, satisfy}
\ben \boxed{ \beta(g) \le  \frac{16 \pi}{3}  \pi  E\,. \label{conj3} }
\een 
We find that this is satisfied in all the cases we have tested.

   Based on an investigation of various higher-dimensional black hole 
examples, we find that an analogue of conjecture 2 in (\ref{conj2}) holds in
all cases.  Thus we propose

\medskip
\noindent
{\bf Conjecture 4}: {\it The length $\ell(g)$ of the shortest closed
geodesic, and the energy
$E$ of an apparent horizon, in $D$ spacetime dimensions, satisfy}
\ben\boxed{
\ell(g)^{D-3} \le \frac{32\pi^{D-2} E}{(D-2) {\cal A}_{D-2}}\,,
}\label{conj4}
\een
where ${\cal A}_{D-2}$ is the volume  of the standard
round $(D-2)$-sphere of unit radius.  Note that in five dimensions,
conjecture 4 does not follow from conjecture 3.\footnote{In fact 
for any metric $g$ on $T^2$ Loewner has shown that  
for the shortest  non-null homotopic curve 
\ben
\ell(T^2,g) \le  \frac{2^\half }{3^{\frac{1}{4}}  } \,\sqrt{ A(T^2,g) } \,.
\een
}

  The results in four dimensions described earlier strongly suggest that
equation (\ref{pu}) holds for all apparent horizons, with or without an
antipodal symmetry. There is no analogue of Pu's theorem in higher dimensions.
Nevertheless our calculations suggest the validity of 

\medskip
\noindent
{\bf Conjecture 5}: {\it The length $\ell(g)$ of the shortest closed geodesic,
and the $(D-2)$-volume $A$ of an apparent horizon, in $D$ spacetime 
dimensions, satisfy, at least in even dimensions,}
\ben\boxed{
\bigl( \frac{\ell(g)}{2 \pi} \bigr )^{D-2} \le \frac{A}{{\cal A}_{D-2} } \,.}
\label{conj5}
\een

  We have verified that conjectures 4 and 5 are both satisfied for
Kerr-AdS black holes in all even dimensions.  We also find that
conjecture 4 is satisfied for Kerr-AdS black holes in all
odd dimensions.  We have so far been unable to find a suitable bound for
$\ell(g)^{D-2}/A$ that would support conjecture 5 in odd dimensions.
 
Note that in five spacetime dimensions the Penrose or Isoperimetric
 Inequality for black holes  is
\cite{Gibbons2,Bray}
\ben
A \le 2 \pi ^2 \bigl(\frac{ 8E }{3 \pi} \bigr) ^{\frac{3}{2}}\,,
\een
but since, even if the metric admits an antipodal map,
there appears  to be no useful general inequality for $ \frac{l^3(g)}{
 A(g)}$
\cite{Croke1,Croke2}, 
this does not give us useful information about conjecture 4.

   In $D$ spacetime  dimensions the known exact  black hole solutions 
admit foliations of the $S^{D-2}$ horizons 
by $T^{[\frac{D-1}{2}]}$ which have co-dimension larger than one if $D\ge 6$. 
Thus
they cannot be used as ``hyperhoops.''  However, 
in the case of rotating black holes with a single
non-vanishing rotation parameter we are able to construct a foliation of
the horizon by leaves with topology $S^1 \times S^{D-4}$. This allows
us to define the  Birkhoff  invariant in terms of the  $(D-3)$-volume of
these ``hyperhoops,'' suggesting 

\medskip
\noindent
{\bf Conjecture 6}: {\it For sweepouts by $S^1\times S^{D-4}$ hyperhoops,
the Birkhoff invariant $\beta(g)$ and the energy
$E$ of an apparent horizon, in dimensions $D=2N+1$,  satisfy}
\ben \boxed{
\beta(g)  \le  \fft{32\pi}{(2N-1)}\, (N-1)^{\ft12(N+1)}\, 
N^{-\ft12 N}\,E \,. }\label{conj6}
\een
We have verified this conjecture for all odd-dimensional Kerr-AdS black holes
with a single non-vanishing rotation parameter, with or without an
external magnetic field.

\section{Birkhoff Bound in Four Spacetime Dimensions}

   In this section we shall test conjecture 1, given in (\ref{conj1}), 
for three explicitly-known classes of black holes in four dimensions.
To begin with, we consider the general 4-charged rotating black holes
of ${\cal N}=8$ ungauged supergravity.  These are asymptotically flat, and
generalise the examples considered in \cite{Gibbons}, which were restricted
to the case of pairwise-equal charges.  The second class we shall consider
is rotating asymptotically-AdS black holes.  These are solutions of 
${\cal N}=8$ gauged supergravity, with pairwise-equal charges.  Finally, 
we shall consider solutions of Einstein-Maxwell theory in which a neutral
black hole is immersed in a Melvin-type magnetic field.  In all cases, 
we find an upper bound for the Birkhoff invariant $\beta(g)$, which is
at most equal to the upper bound given by conjecture 1.

\subsection{Four-dimensional asymptotically-flat black holes}

These 4-charge solutions in ungauged 
${\cal N}=8$ supergravity were obtained in \cite{cvetyoum4}, and
a convenient expression for them can be found in \cite{chcvlupo4}:
\bea
ds_4^2 &=& -\fft{\rho^2-2mr}{W}\, (dt+Bd\phi)^2 + W\Big(\fft{dr^2}{\Delta} +
  d\theta^2 + \fft{\Delta\, \sin^2\theta}{\rho^2-2mr}\, d\phi^2\Big) \,,
\label{4charge}\\
B &=& \fft{2ma (r c_{1234} -(r-2m) s_{1234})\, \sin^2\theta}{
                     \rho^2-2mr}\,,\nn\\
W^2&=& r_1 r_2 r_3 r_4 + a^4\cos^4\theta + \nn\\
&&a^2[2r^2 + 2mr \sum_i s_i^2 
  + 8m^2 c_{1234} s_{1234} - 4m^2 (s_{123}^2 + s_{124}^2 + s_{134}^2 +
  s_{234}^2 + 2 s_{1234}^2)]\cos^2\theta\,,\nn\\
\Delta &=& r^2-2mr + a^2\,,\qquad \rho^2=r^2+a^2\cos^2\theta\,,\nn\\
r_i &=& r+2m s_i^2\,,\qquad s_{i_1\cdots i_n}= s_{i_1}\cdots s_{i_n}\,,
\qquad c_{i_1\cdots i_n} = c_{i_1}\cdots c_{i_n}\,,\nn
\eea
where here, and throughout the paper, we use the abbreviations
\be
s_i= \sinh\delta_i\,,\qquad c_i=\cosh\delta_i\,.
\ee
The metric (\ref{4charge}) depends on the mass parameter $m$, the rotation
parameter $a$, and the four charge ``boost'' parameters $\delta_i$.  In
order to avoid the unnecessary manipulation of square roots, it is 
convenient to use $r_+$, the radius of the outer 
horizon, rather than $m$, in the parametrization.  
Thus we have $m=(r_+^2+a^2)/(2r_+)$.  Special
cases include the 
Kerr metric when $s_i=0$; the Schwarzschild metric if additionally 
$a=0$; the Kerr-Newman metric if $s_i=s$; and the Reissner-Nordstr\"om metric
if additionally $a=0$.

   We shall bound the Birkhoff invariant $\beta(g)$ from above by
considering a foliation by circles $\theta=$constant. 
   It is easily seen that on the horizon, $g_{\phi\phi}$ attains its maximum
value at $\theta=\ft12\pi$, and so setting $f=\theta$ in (\ref{betagf}) we
have
\be
\beta(g) \le \beta(g;\theta)
= \fft{2\pi\, (r_+^2+a^2)\Big(r_+^2 \,\prod_i c_i+ a^2\prod_i s_i\Big)}{
        r_+\, \prod_j[r_+^2\, c_i^2 + a^2 s_i^2]^{1/4}}\,.\label{4hoopL}
\ee
The mass of the black hole is given by \cite{cvetyoum4}
\be
E = \ft14 m \sum_i (c_i^2+s_i^2)\,,
\ee
and therefore conjecture 1 will be satisfied if
\be
\sum_i(c_i^2+s_i^2) - \fft{4[\prod_i c_i + \tilde a^2\, \prod_i s_i]}{
         \prod_j[c_j^2+ \tilde a^2\, s_j^2]^{1/4}}\ \ge \ 0\,,\label{4q1}
\ee
where we have defined $\tilde a \equiv a/r_+$.  We can easily establish
that (\ref{4q1}) is satisfied, by observing that
\be
\sum_i c_i^2 - \fft{4\prod_i c_i}{\prod_j[c_j^2+ \tilde a^2\, s_j^2]^{1/4}}
\ \ge\ \sum_i c_i^2 - \fft{4\prod_i c_i}{\prod_jc_j^{1/2}}=
 \sum_i c_i^2 -4\Big(\prod_i c_i\Big)^{1/2}\ \ge\ 0\,,
\ee
and that
\be
\sum_i s_i^2 -  \fft{4\tilde a^2\,  \prod_i s_i}{\prod_j[c_j^2+ 
    \tilde a^2\, s_j^2]^{1/4}} \ \ge\  
\sum_i s_i^2-  \fft{4\tilde a^2\,  \prod_i s_i}{
 \prod_j[\tilde a^2\, s_j^2]^{1/4}}  =\sum_i s_i^2-4 \Big(\prod_i s_i\Big)^{1/2}
\ \ge\ 0\,,
\ee
where in each case the final inequality follows by using the standard
relation between the geometric and arithmetic mean of non-negative quantities
$x_i$:
\be
\fft1{n}\, \sum_{i=1}^n x_i \ \ge\  \Big(\prod_{i=1}^n x_i\Big)^{1/n}\,.
\label{geoari}
\ee

\subsection{Four-dimensional asymptotically-AdS black holes}

These solutions
for rotating black holes with pairwise-equal charges in ${\cal N}=8$
gauged supergravity, which were obtained in \cite{chcvlupo4}, have the metric
\bea
ds_4^2 &=& -\fft{\Delta_r}{W}\, \Big(dt - a\, \sin^2\theta\,
\fft{d\phi}{\Xi}\Big)^2 + W\,  \Big( \fft{dr^2}{\Delta_r} +
\fft{d\theta^2}{\Delta_\theta} \Big) 
   + \fft{\Delta_\theta\, \sin^2\theta}{W}\, \Big[a\, dt - (r_1 r_2 +
   a^2)\fft{d\phi}{\Xi}\Big]^2\,,\nn\\
\Delta_r &=& r^2 + a^2 - 2m\, r + g^2 \, r_1\, r_2\, (r_1\, r_2 +
a^2)\,,\nn\\
\Delta_\theta &=& 1 - g^2\, a^2\, \cos^2\theta\,,\qquad
W=r_1\, r_2 + a^2 \cos^2\theta\,,\label{4met2}\\
r_1&=& r+ q_1\,,\qquad r_2= r+ q_2\,,\qquad \Xi=1-a^2 g^2\,.\nn
\eea
The physical charges are proportional to $\sqrt{q_i(q_i+2m)}$, and
reality of the solution implies that the charge parameters $q_1$ and 
$q_2$ should be taken to be non-negative. 
Special cases include the Kerr-Newman-AdS metric if $q_1=q_2$, 
and the Kerr-AdS metric if $q_1=q_2=0$.

The maximum value of $g_{\phi\phi}$ on the horizon 
is attained at $\theta=\ft12\pi$, 
and it is easily seen that
\be
\beta(g)\le \beta(g;\theta)
 = \fft{2\pi\, [(r_+ + q_1)(r_+ + q_2)+a^2]}{\Xi\, (r_+ +q_1)^{1/2}\, 
     (r_+ + q_2)^{1/2}}\,,
\ee
where $r_+$, the radius of the horizon, is the largest root of 
 $\Delta_r(r)=0$.  

  The mass of the black hole is given by \cite{timemachine}
\be
E= \fft{2m+q_1+q_2}{2\Xi^2}\,,
\ee
and so conjecture 1 is satisfied if
\bea
&&1+ \tilde q_1 + \tilde q_2 - (1+ \tilde q_1)^{1/2} \, 
  (1+\tilde q_2)^{1/2} + \tilde a^2[1 - (1+\tilde q_1)^{-1/2} \,
(1+\tilde q_2)^{-1/2}] \label{pairbound}\\
&&+ \tilde g^2[(1+\tilde q_1)(1+\tilde q_2) +\tilde a^2][
(1+\tilde q_1)(1+\tilde q_2) +\tilde a^2 (1+ \tilde q_1)^{-1/2} \,
  (1+\tilde q_2)^{-1/2}]\ \ge\ 0\,,\nn
\eea
where we have defined the dimensionless quantities
 $\tilde a=r/r_+$, $\tilde q_i=q_i/r_+$ and
$\tilde g= g r_+$.  Clearly the terms in (\ref{pairbound}) proportional
to $\tilde g^2$ are always positive, as are the bracketed terms with 
the $\tilde a^2$
prefactor.  The positivity of the remaining terms can be seen easily by
squaring:
\be
(1+ \tilde q_1 + \tilde q_2)^2 -(1+\tilde q_1)(1+\tilde q_2)=
  \tilde q_1 + \tilde q_2 + \tilde q_1^2+ \tilde q_2^2+ \tilde q_1 
\tilde q_2\ \ge\ 0\,,
\ee
thus establishing that conjecture 1 is satisfied by 
these metrics.

\subsection{Four-dimensional asymptotically-Melvin black holes}

These were first constructed using a Harrison transformation
in Einstein-Maxwell theory,
by   Ernst \cite{Ernst} and in an explicit form by 
Ernst and Wild \cite{ErnstWild}.
In what follows,
we shall see that, perhaps surprisingly,  conjecture 1 extends 
to asymptotically Melvin solutions, at least in the non-rotating case.
The metric is 
\ben
ds_4^2 = F^2 \Bigl\{ -\bigl (1-{2m \over r}\bigr ) dt^2 +  
{ dr^2   \over 1-{2 m \over r} }    + r^2 d\theta^2  \Bigr\}
+ {r^2 \sin^2\theta \over F^2 }  d\phi^2 \,,  
\label{Ernst}
\een
with 
\ben
F = 1 + {B^2 \over 4} r^2 \sin^2\theta\,,    
\een
where $B$ is the applied magnetic field.
If $m=0$ we get the Melvin solution, whilst if instead $B=0$ we get the
Schwarzschild solution. The energy with respect
to the Melvin background is given simply by \cite{Radu}
\be
E=m\,,
\ee
and the horizon, which is located at
\ben
r=2m\,,
\een 
has the metric
\ben
ds^2 = 4 m^2 \Bigl \{
 (1+ \gamma^2 \sin^2\theta)^2  d\theta^2 + {\sin^2\theta \over 
(1+ \gamma^2 \sin^2\theta )^2 }  d\phi^2  \Bigr \} \,,   
\een
with 
\ben
\gamma  = m\, |B| \,.
\een
The area $A$ and temperature $T$ of the horizon
are
\ben
A= 16 \pi m^2 \,,\qquad T= {1 \over 8 \pi m} \,.
\een
Remarkably, these are the same as in the absence of the magnetic field
\cite{Radu}.

   We have
\ben
\beta(g) \le \sup_\theta \, \frac{4\pi\, E\sin \theta}{1+ \gamma^2
  \sin^2\theta} 
\een
If $\gamma \le 1$, the circumference  $C(\theta)$ has a single maximum
at the equator $\theta =\frac{\pi}{2} $, the maximum value being
$\frac{4  \pi E }{1+\gamma^2}  \le 4 \pi E$. If $\gamma \ge 1$, the horizon is
  dumb-bell shaped
and has two maxima with  $\gamma \sin \theta =1$, the maximum value
being $\frac{2E\pi}{ \gamma} < 4 \pi E$. Thus conjecture 1 is always
satisfied for this metric. 
More about the geometry of the horizon may be found in the appendix.

   The solutions for a black hole immersed in a
magnetic field in Einstein-Maxwell-Dilaton theory
have been given by Yazadjiev \cite{Yazadjiev}. He gives results   
in higher dimensions also, but here we quote the result just for $D=4$. 
The main change is that $F$ in (\ref{Ernst}) is replaced by
$F^{1 \over 1 + \alpha^2}$ where $\alpha$ is the dilaton
coupling constant. The area, surface gravity and mass of the
solution are independent of $\alpha$,  as is the location of the horizon.
The horizon metric is
\ben
ds^2 = 4 E^2 \Bigl \{
 (1+ \gamma^2 \sin^2\theta )^{2 \over 1+ \alpha^2}  
 d\theta^2 + {\sin^2\theta \over 
(1+ \gamma^2 \sin^2\theta)^{2 \over 1+\alpha^2}}  d\phi^2\Bigr \} \,,   
\een
The circumference has a single maximum at $\theta = \frac{\pi}{2}$ 
as long as $\gamma ^2 \le \frac{1+\alpha ^2}{1-\alpha ^2}$,
otherwise it has two maxima when $\sin ^2 \theta =
 \frac{1+\alpha ^2} {\gamma (1-\alpha ^2)} $. 
In all cases $\beta \le 4 \pi E$.

\section{Birkhoff Bound in Five Spacetime Dimensions}

In this section, we consider $S^1\times S^1$ sweepouts.
In this case these coincide with both the $S^1 \times S^{D-4}$ and the
$T^{[{ D-1 \over 2 }] }$ sweepouts that we discussed in the introduction.
One could instead consider sweepouts by $S^2$, but it seems
that the more useful is by 2-tori.  We shall check conjecture 3, given in
(\ref{conj3}), for two classes of five-dimensional black holes.
The first class consists of rotating 3-charge solutions of maximal 
ungauged supergravity, and the second class consists of charged rotating 
black holes in minimal gauged supergravity.

   For a spherical 3-surface $S=\{S^3,g\}$ and a foliation 
$f: S \rightarrow {\Bbb R}$ with generic
level sets $f^{-1}(c)$ being tori of area $A(c)$ and two singular leaves being
linked circles, we define\footnote{A standard example of such a foliation
is to write the round $S^3$ metric as $ds^2=d\theta^2 + \sin^2\theta\, d\phi^2
+\cos^2\theta\, d\psi^2$.}
\be
\beta(g;f)= \hbox{max}_c A(c)\,,\label{5betagf}
\ee
and 
\be
\beta(g)= \inf_f \beta(g;f)\,.\label{5beta}
\ee
 
\subsection{Five-dimensional asymptotically-flat black holes}

  A convenient expression for the rotating 3-charge asymptotically-flat
solutions found in \cite{cvetyoum5} is given in \cite{chcvlupo5}:
\bea
ds_5^2 &=& (H_1 H_2 H_3)^{1/3}\, (x+y)\, 
 \Big(-\Phi\, (dt + {\cal A})^2 + ds_4^2\Big)\,,\label{kkmet}\\
ds_4^2 &=& \Big( \fft{dx^2}{4 X} + \fft{dy^2}{4Y}\Big) 
    + \fft{U}{G}\, \Big(d\chi - \fft{Z}{U}\, d\sigma\Big)^2 +
     \fft{X Y}{U}\, d\sigma^2\,,\nn\\
H_i &=& 1 + \fft{2m s_i^2}{x+y}\,,\qquad 
   \Phi = \fft{G}{(x+y)^3\, H_1\, H_2\, H_3}\,,\nn\\
X &=& (x+a^2)(x+b^2)- 2mx \,,\qquad
Y= - (a^2-y)(b^2-y) \,,\nn\\
G &=& (x+y)(x+y-2m)  \,,\qquad
U = y X - x Y \,,\qquad Z = a b (X+Y) \,,\nn\\
{\cal A}&=& \fft{2m c_1 c_2 c_3}{G} [(a^2+b^2-y) d\sigma -a b d\chi]
  -\fft{2m s_1 s_2 s_3}{x+y}\, (ab d\sigma - y d\chi)\,.
\eea
The coordinates $\sigma$ and $\chi$ are related to standard azimuthal
angles $\phi$ and $\psi$ with $2\pi$ periods by
\be
\sigma=\fft{a\phi-b\psi}{a^2-b^2}\,,\qquad 
\chi= \fft{b\phi-a\psi}{a^2-b^2}\,.
\ee
The $x$ and $y$ coordinates are related to standard radial and latitude
coordinates by
\be
x=r^2\,,\qquad y= a^2\cos^2\theta + b^2\sin^2\theta\,.
\ee

   It is straightforward to see that the area of an $S^1\times S^1$ sweepout
on the horizon at a fixed value of $\theta$ is given by
\be
A(\theta) = \fft{(r^2 c_1 c_2 c_3 + a b s_1 s_2 s_3)
  (r^2+a^2)(r^2+b^2)\sin\theta\, \cos\theta}{r^3\rho\, (H_1 H_2 H_3)^{1/6}}\,,
\ee
evaluated at $r=r_+$, the largest root of $X(r^2)=0$,
where
\be
\rho^2 = x+y=r^2 + a^2\cos^2\theta + b^2 \sin^2\theta\,.
\ee
Unlike the situation in four dimensions, where the $S^1$ sweepout has its
greatest length at the midpoint of the range of the latitude coordinate, here
the maximum value $\beta(g;\theta)$ 
of the $S^1\times S^1$ sweepout area $A(\theta)$ occurs
at a value of $\theta$ that is a quite complicated function of the
parameters of the solution.  Accordingly, in order to test conjecture 3
in this case we shall work with an appropriate upper bound on 
the sweepout area $A(\theta)$.  
In order to do this, it is convenient to assume, 
without loss of
generality, that the rotation parameters $a$ and $b$ are ordered
such that
\be
a^2\ \ge \ b^2\,.\label{abbound}
\ee
Since
\be
\rho^2 \ \ge\ r^2+b^2\,,\qquad \sin\theta\, \cos\theta \ \le\  \ft12\,,
\ee
we shall have
\be
\beta(g;\theta) \ \le\ \hbox{max}_\theta A(\theta) \le  \fft{2\pi^2 
                   (r_+^2 c_1 c_2 c_3 + ab s_1 s_2 s_3)
  (r_+^2+a^2)(r_+^2+b^2)^{1/2}}{
    r_+^2\,\prod_i (r_+^2\, c_i^2 + a^2s_i^2)^{1/6}}\,.
\ee

  The mass of these black hole solutions is given by \cite{cvetyoum4}
\be
E= \ft14 m\pi\, \sum_i(c_i^2 + s_i^2)\,,
\ee
Conjecture 3, given in (\ref{conj3}), is therefore
satisfied if
\be
\sum_i(c_i^2 + s_i^2) -\fft{3\Big(\prod_i c_i + \tilde a\tilde b \prod_i s_i
         \Big)}{(1+\tilde b^2)^{1/2}\, \prod_j (c_j^2 +\tilde a^2 s_j^2)^{1/6}}
\ \ge\  0\,,\label{5test}
\ee
where we have defined the dimensionless parameters $\tilde a=a/r_+$ and
$\tilde b=b/r_+$.  We can assume that $\tilde a \tilde b$ is positive,
since if it were negative the inequality would be more easily satisfied.

   Clearly we have the inequalities
\be
\sum_i c_i^2 - \fft{3\prod_i c_i}{(1+\tilde b^2)^{1/2}\,
  \prod_j (c_j^2 +\tilde a^2 s_j^2)^{1/6}} \ \ge\ 
  \sum_i c_i^2 - \fft{3\prod_i c_i}{(\prod_j c_j^2)^{1/6}} = 
    \sum_i c_i^2-  3\Big(\prod_i c_i\Big)^{2/3} \ \ge \ 0
\ee
and 
\bea
\sum_i s_i^2 -\fft{3 \tilde a\tilde b \prod_i s_i}{
  (1+\tilde b^2)^{1/2}\, \prod_j (c_j^2 +\tilde a^2 s_j^2)^{1/6}} 
&\ge& \sum_i s_i^2 - \fft{3 \tilde a\tilde b \prod_i s_i}{
   (1+\tilde b^2)^{1/2}\, \prod_j(\tilde a^2 s_j^2)^{1/6}}
\nn\\
&=&\sum_i s_i^2 - \fft{3\tilde b\Big(\prod_i s_i\Big)^{2/3}}{
            (1+\tilde b^2)^{1/2}} \ \ge \ 0\,,
\eea
where in each case we have used (\ref{geoari}) in the final step (together
with $\tilde b/(1+\tilde b^2)^{1/2}\ \le \ 1$ in the second case). 
Thus we see that the inequality (\ref{5test}) holds, and so the
five-dimensional 3-charge rotating black holes are indeed consistent with
conjecture 3.

\subsection{Five-dimensional asymptotically-AdS black holes}

   The metric and gauge potential for this solution of minimal gauged
supergravity are given by \cite{chcvlupod5} 
\bea
ds^2 &=& -\fft{\Delta_\theta\, [(1+g^2 r^2)\rho^2 dt + 2q \nu]
\, dt}{\Xi_a\, \Xi_b \, \rho^2} + \fft{2q\, \nu\omega}{\rho^2}
+ \fft{f}{\rho^4}\Big(\fft{\Delta_\theta \, dt}{\Xi_a\Xi_b} -
\omega\Big)^2 + \fft{\rho^2 dr^2}{\Delta_r} \nn\\
&&+
\fft{\rho^2 d\theta^2}{\Delta_\theta}
+ \fft{r^2+a^2}{\Xi_a}\sin^2\theta d\phi^2 + 
      \fft{r^2+b^2}{\Xi_b} \cos^2\theta d\psi^2\,,\label{5met}\\
A &=& \fft{\sqrt3 q}{\rho^2}\,
         \Big(\fft{\Delta_\theta\, dt}{\Xi_a\, \Xi_b} 
       - \omega\Big)\,,\label{gaugepot}
\eea
where
\bea
\nu &=& b\sin^2\theta d\phi + a\cos^2\theta d\psi\,,\qquad
\omega = a\sin^2\theta \fft{d\phi}{\Xi_a} + 
              b\cos^2\theta \fft{d\psi}{\Xi_b}\,,\nn\\
\Delta_\theta &=& 1 - a^2 g^2 \cos^2\theta -
b^2 g^2 \sin^2\theta\,,\qquad f= 2 m \rho^2 - q^2 + 2 a b q g^2 \rho^2\,,\nn\\
\Delta_r &=& \fft{(r^2+a^2)(r^2+b^2)(1+g^2 r^2) + q^2 +2ab q}{r^2} - 2m 
\,,\nn\\
\rho^2 &=& r^2 + a^2 \cos^2\theta + b^2 \sin^2\theta\,,\qquad
\Xi_a = 1-a^2 g^2\,,\quad \Xi_b = 1-b^2 g^2\,,
\eea

   The determinant of the two-dimensional sub-metric spanned by $d\phi$
and $d\psi$ is given on the horizon at $r=r_+$ by
\be
\sqrt{\det(Z_{ij})} = \fft{\sqrt{\Delta_\theta}\, \sin\theta\cos\theta}{
       \Xi_a\, \Xi_b\, r_+\, \rho_+}\, [(r_+^2+a^2)(r_+^2+b^2)+ abq]\,,
\ee
where $r_+$ is the largest root of $\Delta_r(r)=0$, and $\rho_+$ means $\rho$
evaluated with $r=r_+$.  It is convenient to assume, without loss of 
generality, that $a^2\ge b^2$, and so although it is not easy to give
the exact expression for $\sqrt{\det(Z_{ij})}$ maximized over $\theta$,
we may use the inequalities
\be
\rho^2\ge r^2+ b^2\,,\qquad \Delta_\theta \le 1\,,\qquad
\sin\theta\cos\theta\le \ft12
\ee
in order to obtain the upper bound for the area $A(\theta)$ of the 
$S^1\times S^1$ sweepout:
\be
\beta(g;\theta)  \le \hbox{max} A(\theta) \le 
\fft{2\pi^2\,[ (r_+^2+a^2)(r_+^2+b^2)+ abq] }{
  \Xi_a\, \Xi_b\, r_+\, (r_+^2+b^2)^{1/2}} \,.\label{5gaugebound}
\ee
(Sharper bounds can, of course, be obtained, but this one turns out to
suffice.)

The energy of the rotating charged black hole is given by 
\cite{chcvlupod5}
\be
E= \fft{m\pi(2\Xi_a+2\Xi_b - \Xi_a\, \Xi_b) +
2\pi q a b g^2(\Xi_a+\Xi_b)}{4 \Xi_a^2\, \Xi_b^2}\,.\label{5gaugeen}
\ee
It is convenient to parametrize the metric by $a$, $b$, $q$ and $r_+$,
with $m$ solved for in terms of these, and the gauge coupling $g$, by
using $\Delta(r_+)=0$.  If we then form the dimensionless quantities
\be
\tilde a=\fft{a}{r_+}\,,\quad \tilde b=\fft{b}{r_+}\,,\quad
  \tilde q= \fft{q}{r_+^2}\,,\quad \tilde g= g r_+\,,
\ee
then using (\ref{5gaugebound}) and (\ref{5gaugeen}), conjecture 3  
will be verified for this rotating charged black hole if
\be
(1+\tilde a^2)(1+\tilde b^2)(1+\tilde g^2)+ 2\tilde a\tilde b\tilde q
+ \ft83\tilde a\tilde b\tilde g^2\tilde q + \tilde q^2
-  \Big[(1+\tilde a^2)(1+\tilde b^2)^{1/2} 
  +\fft{\tilde a\tilde b\tilde q}{(1+\tilde b^2)^{1/2}}\Big]\ge0\,.
\label{qlin}
\ee

   It is very easy to see that (\ref{qlin}) is satisfied if $q$ is assumed
to be non-negative.  However, in the parametrization used here $q$ can
take either sign.  We therefore proceed by
completing the square on the terms involving $\tilde q$ in (\ref{qlin}).
Dropping the positive term $(\tilde q+\cdots)^2$ implies that (\ref{qlin})
will be satisfied if the inequality
\be
(1+\tilde a^2+\tilde b^2) -\fft{\tilde a^2\tilde b^2(1-\tilde b^2\tilde g^2)}{
  4(1+\tilde b^2)} -\ft19 \tilde a^2\tilde b^2 \tilde g^2(15+16\tilde g^2)
-\fft{1+\tilde a^2+\tilde b^2 -\ft43 \tilde a^2\tilde b^2\tilde g^2}{
   (1+\tilde b^2)^{1/2}}\ge0\label{5test2}
\ee
holds.  We have already assumed that $a^2\ge b^2$, and we must also
restrict the rotation parameters such that $\Xi_a>0$, $\Xi_b>0$, so we
must require that $\tilde a^2 \tilde g^2<1$.  A convenient 
reparametrization that takes account of these conditions and that 
eliminates the square root in (\ref{5test2}) is to write
\be
\tilde a = \ft12 (d_1-d_1^{-1})\,, \qquad
\tilde b=\ft12 (d_2-d_2^{-1})\,,\qquad 
\tilde g=\tilde a^{-1}\, (z+1)^{-1}\,, 
\ee
and then set
\be
d_1=x+y+1\,,\qquad d_2=y+1\,.
\ee
The parameter space is then spanned by $x$, $y$ and $z$ lying in the
positive octant of ${\Bbb R}^3$.  

    Substituting these definitions into (\ref{5test2})
then shows that conjecture 3 is satisfied if a certain 
multinomial $P(x,y,z)$  
is positive for all positive $x$, $y$ and $z$.  $P(x,y,z)$
has  441 terms, of which 438 form a multinomial $Q(x,y,z)$ whose coefficients
are all strictly positive, plus 3 remaining terms with negative coefficients:
\be
P(x,y,z)= Q(x,y,z) -324 y^{11} -145 y^{12} -6 y^{13}\,.
\ee
In fact 
\bea
P(0,y,0)&=& (1+y)^2\, (2304 + 13824 y + 37376 y^2  + 60160 y^3  + 64320 y^4  
     + 48384 y^5  \label{dodec}\\
&&\qquad \qquad + 26144 y^6  + 9696 y^7  + 1968 y^8  - 64 y^9  - 124 y^{10}
       - 12 y^{11}   + 3 y^{12})\,,\nn
\eea
and since $P(x,y,z)-P(0,y,0)$ has strictly positive coefficients, conjecture
3 will be established if we can show that the dodecadic factor in
(\ref{dodec}) is positive for all positive $y$.  This can be shown by 
means of a straightforward application of Sturm's sign-sequence theorem 
\cite{sturm}.  This completes the demonstration that the charged rotating
black hole in five-dimensional minimal gauged supergravity satisfies
conjecture 3 for $S^1\times S^1$ sweepouts.

\section{Closed Geodesic Bounds}

   In this section, our aim is to test conjectures 4 and 5 for Kerr-AdS
black holes in arbitrary dimensions.  In order to do so, we need a bound
on the length $\ell(g)$ of the shortest closed non-trivial geodesic.  By
a theorem of Lyusternik and Fet, every compact Riemannian manifold
admits at least one nontrivial closed geodesic \cite{lyfe}.  Moreover, it
is a long-standing conjecture that there exist infinitely many nontrivial
closed geodesics on every compact Riemannian manifold \cite{yau2}.  
For any metric on the 3-sphere, it has been shown that there are at least
two geometrically distinct closed geodesics \cite{lodu}.

  From the results quoted above, we 
may assume that the horizons discussed in this paper 
admit a closed geodesic of shortest length.
In fact, we may exhibit explicitly at least 
$\ft14 (D-1)^2$ closed geodesics in 
the odd-dimensional Kerr-AdS metrics, and $\ft14(D^2-4)$ in the 
even-dimensional cases.

   In four spacetime dimensions, Pu's theorem \cite{Pu} gives an upper 
bound for the length of the shortest closed geodesic in terms of the area
of the horizon, provided that the horizon admits an antipodal map.  All of
the horizons we consider in this paper do in fact admit an antipodal
symmetry, but it appears that there is no higher-dimensional generalisation
of Pu's theorem.  Nevertheless, it is still possible to give an upper
bound to the length of the shortest closed geodesic, by bounding the
length of any curve joining a point and its antipode.  This will provide
a bound on what is called the ``systole,'' which is defined as the least 
length of any homotopically nontrivial curve on the quotient of the
horizon by the antipodal map \cite{berger}. In fact, in what follows we 
shall estimate
the systole by finding closed geodesics that pass through pairs of
antipodal points.

\subsection{Asymptotically-AdS black holes in higher dimensions} 

    The general Kerr-AdS metrics in arbitrary dimension $D$
were obtained in \cite{gilupapo1,gilupapo2}.  They have $N\equiv [(D-1)/2]$
independent rotation parameters $a_i$ in $N$ orthogonal 2-planes. 
We have $D=2N+1$ when $D$ is odd, and $D=2N+2$ when $D$ is even. Defining
$\ep\equiv (D-1)$ mod 2, so that $D=2N+1+\ep$, the metrics can be 
described by introducing $N$
azimuthal angles $\phi_i$, and $(N+\ep)$ ``direction cosines'' $\mu_i$
obeying the constraint
\be
\sum_{i=1}^{N+\ep} \mu_i^2 =1\,.\label{muconstraint}
\ee
In Boyer-Linquist coordinates, the metrics are given by 
\cite{gilupapo1,gilupapo2}
\bea
ds^2 &=& - W\, (1 + g^2 r^2)\, dt^2
 + \fft{2m}{U}\Bigl(W\,dt
 - \sum_{i=1}^N \fft{a_i\, \mu_i^2\, d\phi_i}
  {\Xi_i }\Bigr)^2
 + \sum_{i=1}^N \fft{r^2 + a_i^2}{\Xi_i}\,\mu_i^2\,
    d\phi_i^2 \nn\\
&&
 + \fft{U\, dr^2}{V-2m}
 + \sum_{i=1}^{N+\ep} \fft{r^2 + a_i^2}{\Xi_i}\, d\mu_i^2
 - \fft{g^2}{W\, (1 + g^2\, r^2)}
    \Bigl( \sum_{i=1}^{N+\ep} \fft{r^2 + a_i^2}{\Xi_i}
    \, \mu_i\, d\mu_i\Bigr)^2 \,,\label{bl}
\eea
where
\bea
W &\equiv& \sum_{i=1}^{N+\ep} \fft{\mu_i^2}{\Xi_i}\,,\qquad
U \equiv  r^{\ep}\, \sum_{i=1}^{N+\ep} \fft{\mu_i^2}{r^2 + a_i^2}\,
\prod_{j=1}^N (r^2 + a_j^2)\,,\label{uwline}\\
V &\equiv& r^{\ep-2}\, (1 +g^2 r^2 )\,
   \prod_{i=1}^N (r^2 + a_i^2)\,,\qquad \Xi_i\equiv 1 - g^2\, a_i^2\,.
\label{uvw}
\eea
They satisfy $R_{\mu\nu}=-(D-1)\,g^2 \, g_{\mu\nu}$.  The horizon is
located at $r=r_+$, where $r_+$ is the largest root of $V(r)=2m$.
The induced metric on the horizon is obtained by setting $r=r_+$ and
$t=\,$constant in (\ref{bl}).

   For our purposes it will prove more illuminating to introduce $2N$
Cartesian coordinates $(x_i,y_i)$ and, in even spacetime dimensions, where
$\ep=1$, an additional coordinate $z$, such that
\be
x_i+i y_i = \mu_i\, e^{i\phi_i}\,,\qquad z= \ep \mu_{N+\ep}\,.
\ee
The constraint (\ref{muconstraint}) becomes
\be
\sum_{i=1}^N (x_i^2+y_i^2) + z^2 =1\,,
\ee
which defines a round hypersphere in ${\Bbb E}^{2N+\ep}$. One has, for each 
$i$,
\bea
\mu_i d\mu_i &=& x_i dx_i + y_i dy_i \,,
 \qquad \ep \mu_{N+\ep} d\mu_{N+\ep} =z dz\,,\nn\\
\mu_i^2 d\phi_i &=& x_i dy_i - y_i dx_i\,,\nn\\
d\mu_i^2 + \mu_i^2 d\phi_i^2 &=& dx_i^2 + dy_i^2\,.
\eea

   On the horizon, we have the following commuting ${\Bbb Z}_2$ isometries:
\bea
A:\qquad\qquad x_i &\longrightarrow&  -x_i\,,\nn\\
B:\qquad\qquad y_i &\longrightarrow&  -y_i\,,\nn\\
C_i:\qquad (x_i,y_i) &\longrightarrow& (-x_i,-y_i) \qquad 
\hbox{for each } i\,,\nn \\  
D:\qquad\qquad z &\longrightarrow& -z \,.\label{ABCD}
\eea
In each case, those coordinates that are not specified are left unchanged
by the map.  For the maps $A$ and $B$, all $N$ of the $x_i$ or $y_i$ 
coordinates undergo a sign reversal.  For the map $C_i$, only the $x_i$
and $y_i$ coordinates for the specified value of $i$ undergo a sign reversal.
The product $ABD$ is the antipodal map
\ben
(x_i,y_i, z) \rightarrow (-x_i,-y_i, -z)\,.
\een
The fixed-point sets of any product of $A$, $B$, $C_i$, $D$ are 
totally-geodesic submanifolds.
If the fixed-point set is one-dimensional it is a geodesic; if it is
two-dimensional, it is
a minimal (strictly, extremal) 2-surface;  etc.
All of the isometries lift to the whole spacetime
provided that either $t$ is unchanged or $t \rightarrow -t$ as appropriate.

   Using these facts, one easily shows that the following circles are geodesic:
\bea
D=2N+2:&& x_i^2 + z^2=1\,,\qquad 1\le i\le N\,,\nn\\
&& y_i^2 + z^2=1\,,\qquad 1\le i\le N\,,\nn\\
&& x_i^2 + x_j^2=1\,,\qquad  1\le i <j\le N\,,\nn\\
&& y_i^2 + y_j^2=1\,,\qquad  1\le i <j\le N\,,\nn\\
&& x_i^2 + y_i^2=1\,,\qquad 1\le i\le N\,,\nn\\
&&\nn\\
D=2N+1:&& x_i^2 + x_j^2=1\,,\qquad  1\le i <j\le N\,,\nn\\
&& y_i^2 + y_j^2=1\,,\qquad  1\le i <j\le N\,,\nn\\
&& x_i^2 + y_i^2=1\,,\qquad 1\le i\le N\,.
\eea
Acting with the isometry group $T^N$, which corresponds to rotations in
each of the $(x_i,y_i)$ planes, one obtains continuous families of
such circular geodesics.  Thus, for generic values of the $a_i$
rotation parameters, there are $\ft12 N(N-1) + 2N=\ft18 (D-2)(D+4)$
classes of geometrically distinct closed geodesics in even dimensions, 
and $\ft12 N(N-1)+N= \ft18 (D^2-1)$ in odd dimensions.  In what follows,
we shall select the closed geodesics that give the optimal estimate
for $\ell(g)$.

\subsubsection{$D=2N+2$ dimensions}

   Let us assume, without
loss of generality, that the rotation parameters are ordered so that
\be
a_1^2\le a_2^2\le a_3^2\cdots \le a_N^2\,.\label{ordering}
\ee
Points on the horizon of the form $x_1^2 + z^2=1$,
with all other $x_i$ and all $y_i$ vanishing, are invariant under the product
$B\prod_{i\ge2} C_i$ of the ${\Bbb Z}_2$ isometries defined in (\ref{ABCD}).
The curve defined by $x_1^2+z^2=1$, which may be parameterised by
\be
x_1=\sin\psi\,,\qquad z=\cos\psi\,,\label{curve}
\ee
for $0\le\psi\le 2\pi$, is therefore a closed geodesic.  Points $\psi$ and
$\psi+\pi$ on the curve are antipodal.

   We see from (\ref{bl}) that the length of this geodesic, $L=\int ds$, is
bounded from above by taking
\be
ds^2\le \Big[r_+^2 \cos^2\psi + \fft{r_+^2+a_1^2}{\Xi_1}\, \sin^2\psi
   \Big]\, d\psi^2\,,
\ee
with equality if $g=0$.
Clearly we may then obtain the bound
\be
ds^2\le \fft{r_+^2+a_1^2}{\Xi_1}\, d\psi^2\,,
\ee
and hence, in view of (\ref{ordering}), 
 the length $L$ of the shortest closed geodesic of this
type is bounded by
\be
L \le \fft{2\pi \,(r_+^2+a_1^2)^{1/2}}{\Xi_1^{1/2}}\,.\label{evenhoop}
\ee
(Note that equality holds in the Schwarzschild limit.)  

The area of the horizon in $D=2N+2$ dimensions is given by 
\ben
A = {\cal A}_{D-2} \prod _i\frac{r_+^2 + a_i^2}{\Xi _i} \,,
\een
where
\be
{\cal A}_{D-2} = \fft{2\pi^{(D-1)/2}}{\Gamma[(D-1)/2]}\label{sphvol}
\ee
is the volume of the unit $(D-2)$-sphere.  It then follows that 
\ben
\bigl ( \frac{l}{2 \pi} \bigr )^{D-2} \le \Bigl( \frac{A}{{\cal A}_{D-2}} 
 \bigr )\,.
\een
Thus conjecture 5, given in (\ref{conj5}), is obeyed in this case.
   
The energy of the Kerr-AdS metric in $D=2N+2$ dimensions is given by
\cite{gibperpop}
\be
E= \fft{m {\cal A}_{D-2}}{4\pi\prod_i \Xi_i}\, \sum_{j=1}^N \fft1{\Xi_j}\,.
\ee
For the Schwarzschild limit we
have $E= m N {\cal A}_{D-2}/(4\pi)$ and $m=\ft12 r_+^{2N-1}$. 
Conjecture 4, given in (\ref{conj4}), is thus 
\be
\fft{8\pi\,  E}{N\, {\cal A}_{D-2}} - \Big(\fft{L}{2\pi}\Big)^{2N-1}\ge 0\,,
\ee
which is saturated in the Schwarzschild limit.

  To test conjecture 4 for the Kerr-Ads metric in $D=2N+2$ dimensions, we
may use the inequality (\ref{evenhoop}), and check to see whether
\be
\fft{(1+g^2 r_+^2)}{N r_+}\, \prod_{i=1}^N \fft{r_+^2+a_i^2}{\Xi_i} \,
  \sum_{j=1}^N \fft1{\Xi_j} - \fft{(r_+^2+a_1^2)^{N-\ft12}}{
  \Xi_1^{N-\ft12}} \ge0\,.
\ee
Reorganizing this as
\be
(1+g^2 r_+^2)\, \Big(1+\fft{a_1^2}{r_+^2}\Big)^{1/2}\, 
  \fft{\prod_{i=1}^N (r_+^2 +a_i^2)}{(r_+^2+a_1^2)^N}\, 
  \Big(\prod_{j=1}^N \fft{\Xi_1}{\Xi_j}\Big)\, \fft1{\sqrt{\Xi_1}} \,
\Big(\fft1{N}\, \sum_{k=1}^N \fft1{\Xi_k}\Big)  -1 \ge0\,,
\ee
we observe that in view of (\ref{ordering}), every factor in the first term 
is greater than or equal to 1, and hence conjecture 4 is 
indeed satisfied by Kerr-AdS in all even dimensions. 

\subsubsection{$D=2N+1$ dimensions}

   The case of odd spacetime dimensions is very similar. The curve 
$x_1^2+x_2^2=1$, with all other $x_i$ and all $y_i$ vanishing, is easily
seen, by arguments similar to those above, to be a closed geodesic. If
\be
x_1=\sin\psi\,,\qquad x_2=\cos\psi\,,
\ee
with $0\le\psi\le 2\pi$, then again points $\psi$ and $\psi+\pi$ are
antipodal.  The length of the closed curve is bounded above by
\be
L\le \int_0^{2\pi} \Big(\fft{r_+^2+a_1^2}{\Xi_1}\,\sin^2\psi +
  \fft{r_+^2+a_2^2}{\Xi_2}\,\cos^2\psi\Big)^{1/2}\, d\psi\,,\label{path1}
\ee
with equality when $g=0$.
If we again assume the rotation parameters are ordered as in (\ref{ordering}),
we obtain the bound
\be
L\le \fft{2\pi(r_+^2+a_2^2)^{1/2}}{\Xi_2^{1/2}}\,.\label{Lodd}
\ee

   The area of the horizon 
in the case of odd spacetime dimensions is given by
\ben
A= \frac{{\cal A}_{D-2}}{r_+} \prod_i\fft{r_+^2 + a_i^2}{\Xi_i}\,.
\een
We find that the bound (\ref{Lodd}) is too weak
to provide support for conjecture 5 in this odd-dimensional case. 

  The mass of the odd-dimensional Kerr-AdS black hole is given by 
\cite{gibperpop}
\be
E= \fft{m\, {\cal A}_{D-2}}{4\pi\prod_i\Xi_i}\,
\Big(\sum_j\fft1{\Xi_j} -\fft12\Big)\,.
\ee
The inequality in conjecture 4, which is saturated in the 
Schwarzschild limit, is then given by
\be
\fft{16\pi\, E}{(2N-1)\, {\cal A}_{D-2}} - 
  \Big(\fft{L_{\rm min}}{2\pi}\Big)^{2N-2}\ge0\,.
\label{oddantibound}
\ee
Substituting the results obtained above, we therefore find that conjecture 4
will be satisfied if
\be
\fft{(1+g^2 r_+^2)}{\Xi_1}\,\fft{(r_+^2+a_1^2)}{r_+^2}\, \, 
\fft{1}{(2N-1)}\, \Big(2\sum_j\fft1{\Xi_j} -1\Big) 
\, \prod_{k=3}^N\Big(\fft{\Xi_2}{\Xi_k}\,\times\, 
\fft{r_+^2+a_k^2}{r_+^2 + a_2^2}\Big) -   1 \ge \ 0\,.
\ee
In view of (\ref{ordering}), we see that conjecture 4 is indeed satisfied
by Kerr-AdS black holes in all odd dimensions.

\section{Sweepouts by Higher-Dimensional Spheres}

  In this section we shall consider more general sweepouts 
by products of spheres $S^p\times S^q$ where $p+q=D-3$.  The definitions of
$\beta(g;f)$ and $\beta(g)$ are completely analogous to those given earlier.

\subsection{$S^1\times S^{D-4}$ sweepouts in $D=2N+1$ Kerr-AdS}

   This can be applied conveniently in the case that all the rotation
parameters are equal.  The metric in $D=2N+1$ dimensions is then given by
\cite{gibperpop}
\be
ds^2 = -\fft{(1+g^2 r^2) dt^2}{\Xi} + \fft{U\, dr^2}{V-2m} +
\fft{r^2+a^2}{\Xi}\, [(d\psi+A)^2 + d\Sigma_{N-1}^2] +
   \fft{2m}{U\Xi^2}\, [dt -a (d\psi+A)]^2\,,\label{oddkads}
\ee
where
\be
U= (r^2+a^2)^{N-1}\,,\qquad V=\fft1{r^2}\, (r^2+a^2)^N\, (1+g^2 r^2)\,,
\qquad \Xi=1-g^2 a^2\,,
\ee
and $A$ is a potential for the K\"ahler form of the Fubini-Study metric
$d\Sigma_{N-1}^2$ on $CP^{N-1}$.  We may write the metric on $CP^{N-1}$
in terms of the Fubini-Study metric on $CP^{N-2}$ as \cite{homapo}
\be
d\Sigma_{N-1}^2 = d\xi^2 +\sin^2\xi\cos^2\xi (d\tau+B)^2 + 
\sin^2\xi d\Sigma_{N-2}^2\,,\label{nested}
\ee
where $B$ is a potential for the K\"ahler form of $d\Sigma_{N-2}^2$.  
The level surfaces of (\ref{nested}) at constant $\xi$ are squashed
$(2N-3)$-spheres, degenerating to a point at $\xi=0$ and to a $CP^{N-2}$
bolt at $\xi=\ft12\pi$.  The volume of the $(2N-3)$-sphere at a given
$\xi$ is
\bea
V_{2N-3} &=& 2\pi\, \sin^{N-1}\xi\, \cos\xi\, \Sigma_{N-2}\,,\nn\\
&=& \sin^{N-1}\xi\, \cos\xi\, \Sigma_{N-2}\, {\cal A}_{2N-3}\,,
\eea
where $\Sigma_{N-2}$ is the volume of the ``unit'' $CP^{N-2}$ metric, and
hence $2\pi \Sigma_{N-2}={\cal A}_{2N-3}$, the volume of the unit round
$(2N-3)$-sphere.  $V_{2N-3}$ attains its maximum volume at 
$\cos\xi=1/\sqrt{N}$, and hence 
\be
V_{2N-3}^{\rm max} =(N-1)^{\ft12(N-1)}\, N^{-\ft12 N}\, {\cal A}_{2N-3}\,.
\ee

    We may now determine the $S^1\times S^{2N-3}$ hyperhoop 
volume,\footnote{In dimensions higher than five, we shall always refer to 
volumes, rather than lengths or areas as we did in four and five dimensions,
when describing codimension-one hyperhoops.} where
the $S^1$ is parameterized by the Hopf fibre coordinate $\psi$ and the
$S^{2N-3}$ is the equatorial sphere obtained above.  In the Kerr-Ads metric
(\ref{oddkads}) we therefore find
\be
V_{\rm hoop} = \fft{2\pi}{r_+}\, \Big(\fft{r_+^2+a^2}{\Xi}\Big)^{N-\ft12}\,
 (N-1)^{\ft12(N-1)}\, N^{-\ft12 N}\, {\cal A}_{2N-3}\,.
\ee
The energy of the Kerr-AdS metric is given by \cite{gibperpop}
\bea
E &=& \fft{m \, (2N-\Xi)\, {\cal A}_{2N-1}}{8\pi\, \Xi^{N+1}}\,,\nn\\
&=& \fft{(2N-\Xi)(r_+^2+a^2)^N\, 
               (1+g^2 r_+^2){\cal A}_{2N-1}}{16\pi\,\Xi^{N+1}}\,.
\eea

   The hyperhoop inequality of conjecture 6 for 
black holes of dimension $D=2N+1$
is given in (\ref{conj6}).  Verifying this for Kerr-AdS requires showing that
\be
\Big(1+\fft{a^2}{r_+^2}\Big)^{1/2}\, (1+g^2 r_+^2) \, \fft1{\sqrt{\Xi}}\, 
 \Big(\fft{2N-\Xi}{2N-1}\Big) -1 \ge0\,.
\ee
Each of the factors in the first term is manifestly greater than or equal to
1, and hence conjecture 6 holds.

\subsection{Sweepouts of $S^{D-2}$ 
 by $S^p \times S^q$ }

There are numerous ways of sweeping out 
 $S^{D-2}$. Among them are sweepouts by products of spheres.
Thus on  the unit round $(D-2)$-sphere, with $D-2= (p+q+1)$,
we may write the metric as
\ben
d \Omega _{p+q+1}^2 = d \theta^2 + \sin^2 \theta \, d\Omega_p^2 + 
\cos^2 \theta \,d \Omega^2_q \,. 
\een
The $(p+q)$-volume of the $S^p \times S^q$ hyperhoop,
\ben
V_{p,q}(\theta)  = \sin^p \theta\, \cos^q \theta \, {\cal A}_p {\cal A}_q 
\,,
\een
is maximized at $\sin \theta = \sqrt{\frac{p}{p+q}}$, $\cos \theta = 
\sqrt{\frac{q}{p+q}}$. 
The maximum value of $V_{p,q}$ is therefore
\ben
V_{p,q} = \bigl( \frac{p}{p+q}\bigr)^{\frac{p}{2}}
 \bigl( \frac{q}{p+q}\bigr)^{\frac{q}{2}} \, 
\frac{
4\pi ^{\frac{p+q+2}{2}}  } 
{\Gamma(\frac{p+1}{2} ) \Gamma(\frac{q+1}{2})}  \,.
\een

In this way we obtain different upper bounds
for the Birkhoff invariant, depending on which 
sweepout we use. Presumably the actual value
of the Birkhoff invariant will also depend on how
we sweep out the sphere. It sometimes happens, because of a suitable symmetry
group,  that one  may define 
sweepouts by products of spheres even if 
the metric on the horizon is not the round one.
In what follows we shall do this for static black holes immersed
in a magnetic field and to rotating black holes with a single
nonvanishing angular velocity.

\subsubsection{Higher-dimensional magnetic fields}

   The metric of a  Tangherlini 
black hole immersed in a magnetic field in $D$ spacetime dimensions 
has been given by Ortaggio \cite{Ortaggio}. The horizon metric is
\ben
ds_H^2 = F^{2 \over D-3}_+ r_+^2 [ \cos^2 \theta d \Omega^2_{D-4}
+ d \theta^2 ] +  F^{-2 }_+ r_+^2 \sin^2 \theta d \phi^2 \,, 
\label{Ortaggio} 
\een
with $F_+ \equiv F(r=r_+)$, 
\ben
F = 1 + {(D-3) \over 2(D-2)} B^2  \rho^2 \,,  
\een
$\rho=  r \sin \theta $, and $d \Omega^2_{D-4}$ 
is the standard round metric on a unit $S^{D-4}$.

The level sets $\theta={\rm constant}$ now  provide a foliation
or  sweepout  of the horizon, whose nonsingular 
leaves have topology $S^\times S^{D-3}$. 
There are two critical level sets, 
$\theta =0$ and $\theta = {\pi \over 2}$.
The first is an $S^{D-4}$, the second an $S^1$.
Note that if $D=5$, the sweepout is by {\it Clifford tori} $S^1\times S^1$.

The $(D-3)$-volume of the hyperhoops with $\theta = {\rm constant}$ is
\ben
V(\theta)= \pi F_+^{-{1 \over D-3}} r_+^{D-3} {\cal A} _{D-4} 
\cos^{D-5}\, \theta  \sin 2\theta  \,.  
\een
Clearly
\ben
V(\theta) \le \pi  r_+^{D-3} {\cal A}_{D-4} \cos^{D-5}\,  \theta  \sin
2\theta  \,.  \label{sweep}
\een
It follows that the magnetic field reduces the
maximum value of $V(\theta)$ below  
the maximum value for $V(\theta)$ on a round sphere for 
this type of sweepout.   
This is given by the value of the right-hand side of (\ref{sweep}) 
when $\cos \theta = \sqrt{\frac{D-5}{D-4} }$. 
This establishes an inequality of the same type as conjecture 6
for this type of sweepout. However it remains unclear     
how this type of sweepout compares with other types of sweepout.

\subsubsection{Non-rotating Einstein-Maxwell-Dilaton 
black holes in higher dimensions}

This was  dealt with by Yazadjiev \cite{Yazadjiev}.
The general results of Ortaggio \cite{Ortaggio} go through with
$F$ in (\ref{Ortaggio} ) replaced by
$F^{{1 \over 1+\alpha^2 }}$,
where $\alpha$ a dilaton coupling constant.
Thus   the  results of the previous section, for which  $\alpha=0$,
still  go through.

\subsubsection{Kerr-AdS  with a single rotation parameter}

    Geometrically this is very similar to the magnetic field case
discussed above.  If the magnetic field vanishes,
the metric on the horizon is 
\ben
ds^2  = \frac{(1- a^2 g^2 \cos^2 \theta) }{(r_+ ^2 + a^2 \cos^2 \theta)}  
\frac{ (r_+^2 + a^2  )^2 }{(1-a^2 g^2)^2  }
  \sin^2 \theta d \phi^2 + r_+^2 \cos^2 \theta\,
d \Omega_{D-4}^2 + \frac{(r_+^2 + a ^2 \cos^2 \theta)}{(1-a^2 g^2
 \cos^2 \theta) }  \, d \theta^2\,,
\een
with $0\le \theta \le {\pi \over 2}$.
Again we have a foliation by $S^1\times S^{D-4}$, and
\ben
V(\theta)=  \pi {\cal A}_{D-4} \,r_+^{D-4} \,  
\sqrt{ \frac{1- a^2 g^2 \cos^2 \theta }{r_+^2 + a^2 \cos^2 \theta} } 
\,\frac{ (r_+^2 + a^2)}{(1-a^2 g^2)} \, \cos^{D-5} \theta
\sin 2 \theta \,.  
\een
The argument now is almost the same as
with the magnetic field, and again conjecture 6 holds for this sweepout.

\subsubsection{Myers-Perry with an applied 
 magnetic field along one non-rotating direction}

  As mentioned above, applying a magnetic field to a rotating
or charged black hole produces quite complicated results owing to
various induction effects. However, if the magnetic
field lies as along a direction (i.e. a two-plane direction) about which
the black hole is not rotating, then   Yazadjiev \cite{Yazadjiev}
has shown that even in Einstein-Maxwell dilaton theory
the metric remains remarkably simple.  
In the case of odd spacetime dimension $D$,
with one rotation parameter (vanishing in the 1-2 plane, say)
and the Maxwell 2-form's having ``legs'' only in the 1-2 direction, then 
the mass, area, location and thermodynamics remain unchanged.
If $ \mu_1, \phi_1$ are the relevant coordinates associated with the  
1-2 plane, so that $a_1=0$ ,  then the horizon metric is given by
eqn  (68) in   \cite{Yazadjiev}:
\bea
ds ^2_H &=& F_+^{- \frac{2}{((D-3) (1+\alpha^2)} } 
\Bigl \{ \sum_{i=2}^{\frac{D-1}{2} }(r_+^2 + a_i^2 ) ( d \mu_i^2 +\mu_i^2 
d \phi_i^2) + \frac{m}{U_+} \Bigl( \sum^{\frac{D-2}{2}}_{i=2} 
a_i \mu_i d\phi \Bigr)^2 + r_+^2 d\mu_1^2 \Bigr \} 
\nonumber \\ 
&+& F_+^{-\frac{2}{1+\alpha^2}} \,r_+^2 \, \mu_1^2 d\phi_1^2\,,\\
U&=& \sum_i\fft{\mu_i^2}{r^2+a_i^2}\, \prod_j(r^2+a_j^2)\,.\nn
\eea 
Now we consider the foliation whose $S^1  \times S^{D-4}$ leaves 
are given by $\mu_1$ = constant. The volume $V(\mu_1, B) $ of
such leaves (where
$B$ is the magnetic field)  satisfies
\ben
V(\mu_1, B) = F_+^{-\frac{1}{(D-3)(1+\alpha^2) }}
V( \mu_1,0) \le V( \mu_1, 0) \,,
\een
and so the magnetic field can only have the effect of 
reducing any upper bound for the Birkhoff invariant.
Thus if conjecture 6 is satisfied in the absence of a magnetic field, it 
will be satisfied in its presence.

\section{Conclusions}

In this paper we have tested some conjectures \cite{Gibbons}  relating   
the geometry of apparent
horizons and their total energy in  four spacetime dimensions, and 
generalized then to 
higher-dimensional spacetimes. 
We expect their validity to depend on a suitable energy condition, but 
this is presumably weaker than the dominant energy condition, since the
latter does not hold in our gauged supergravity examples.
The total energy can be defined in asymptotically-flat, 
asymptotically-AdS and asymptotically-Melvin spacetimes. 
So far we have found support for our conjectures in all even dimensions.
In odd spacetime dimensions we found support for conjectures 4 and 6, 
but we were
unable to make a statement about conjecture 5 (which relates the
ratio of the length of the 
shortest non-trivial geodesic to  to the cube  root of the area). 
The absence of support for conjecture 5 in odd dimensions
is because our upper bound on the length of the shortest nontrivial
closed geodesic is too weak to be decisive. 
If it is in fact a good estimate for $\ell(g)$, then
conjecture 5 would fail in odd dimensions.

  The differences between even and odd dimensions are rather striking, 
and may be related to  other 
differences in the properties
of black holes in even and odd dimensions; for example stability and uniqueness.
For instance, there is growing evidence that  the geometry of 
the horizon plays an important role in determining stability.

Of course failure to find a contradiction to a conjecture
is not a proof, merely ``circumstantial evidence.'' 
However in the course of the investigation
we found that to establish  the necessary inequalities required some 
far from obvious
manipulations.  This, together the number of non-trivial examples, gives 
us some confidence
that the conjectures that have held up so far 
may indeed be true.  It may be possible to give 
some partial proofs
for such configurations as collapsing shells, along the lines of what was 
done in four spacetime dimensions
in \cite{Gibbons}. On the other hand we invite the skeptical reader to 
provide counter-examples 

\section{Acknowledgments}

We would like to thank Thiti Sirithanakorn for 
help during an early part of this investigation. We thank the Mitchell
family for their generous hospitality at Cook's Branch Conservancy during the
completion of this work.  G.W.G. is grateful for hospitality at the University
of Pennsylvania during the early stages of this work.
M.C. is supported in part by DOE grant DE-FG05-95ER40893-A020, 
NSF RTG grant DMS-0636606, the Fay 
R.and Eugene L. Langberg Chair, and the Slovenian Research Agency (ARRS).
The research of C.N.P. is
supported in part by DOE grant DE-FG03-95ER40917.

\appendix

\section{Isometric embedding of the Ernst-Wild horizon in 
${\Bbb E}^3$}

In what follows we give more details of the horizon geometry 
and correct some statements in \cite{WildKerns}.

If the surface can be isometrically embedded as a surface of revolution
in ${\Bbb E} ^3$  then we must have (in units in which  $2E=1$) 
\ben
\rho =  {\sin  \theta \over 
(1+ \gamma ^2 \sin^2\theta)}
\een
and 
\ben
dz ^2 + d \rho ^2 = (1+ \gamma^2 \sin^2\theta )^2  d\theta^2\,,
\een
where $z$, $\rho$ and $\phi$ are cylindrical coordinates for ${\Bbb E}^3$. 
We have
\bea
d \rho &=&  {1-\gamma^2 \sin^2  
\theta \over (1+\gamma^2 \sin^2\theta)^2  } \,\cos\theta  d\theta  \,,\\   
dz & =& { d\theta \over (1+ \gamma^2 \sin^2\theta)^2 }
\sqrt { (1+ \gamma^2 \sin^2\theta )^6 - \cos^2\theta 
(1-\gamma^2\sin^2\theta^2)^2} \,.   
\eea
Now $(1+ \gamma^2 \sin^2\theta)^6 \ge (1- \gamma^2\sin^2\theta)^2$
and $\cos^2\theta \le 1$, and so the argument of the square root
is positive for all $\theta$ and for all $\gamma$. 
{\sl Thus the embedding is global for all $\gamma$}, 
contrary to a statement by  
Wild and Kerns \cite{WildKerns}.

If $\gamma <1$, the surface is convex
circumference $C(\theta)$ of the circular leaves
of the foliation given by $\theta ={\rm constant}$, i.e.
by horizontal planes orthogonal to the axis of symmetry 
is greatest on the equator $\theta = {\pi \over 2}$ where it takes
the value (restoring units),
\ben
C({\pi \over 2}) = {4 \pi E \over 1+ \gamma^2 } \le 4 \pi E\,. 
\een  
 However if $\gamma >1$, then in the interval 
\ben
   \arcsin{1 \over \gamma} < \theta < \pi - \arcsin  {1 \over \gamma} \,,  
\een
which we call the waist region, 
\ben
{d \rho \over d z} <0\,
\een
which means that the surface becomes dumb-bell shaped and hence 
non-convex.
As a consequence the Gauss curvature becomes negative
in a neighbourhood of the equator,  as correctly observed in
\cite{WildKerns}. However that does not preclude a global isometric embedding
into Euclidean 3-space, as we have seen. 

  Recall that Thorne's hoop conjecture was that 
\begin{quote}
{\it Horizons form when and only when a mass $E$ gets
compacted into a region whose circumference in EVERY direction is
$C\le 4 \pi E$.}
\end{quote} \medskip The capitalization  \lq \lq EVERY \rq \rq was 
intended to emphasise the fact that while for the collapse of 
oblate shaped bodies,  the circumferences are all roughly equal,
in the  prolate case, the collapse of a long almost cylindrically 
shaped body whose girth was nevertheless small would not necessarily
produce a horizon.  
However, the  polar circumference of the Schwarzschild-Melvin black hole is
\ben
C_p= 4 E \int _0^\pi (1+ \gamma ^2 \sin ^2 \theta ) \,d \theta 
= 4 \pi E(1+ \half \gamma  ^2) \ge 4 \pi E \,. 
\een
Evidently, the ratio $C_p/(4\pi E)$ may be made arbitrarily large by 
choosing $\gamma$ to be arbitrarily large.  Thus if one were to interpret
$C_p$ as the circumference in directions orthogonal to the axis of
symmetry, then Thorne's conjecture would fail.

\end{document}